\documentclass[%
 reprint,
 superscriptaddress,
 amsmath,amssymb,
 aps,
 prl,
 floatfix,
]{revtex4-2}
\usepackage{float}
\usepackage{graphicx}
\usepackage{dcolumn}
\usepackage{bm}
\usepackage{xcolor}

\usepackage{hyperref}

\usepackage{soul}
\usepackage{comment}
\usepackage{siunitx}
\usepackage[export]{adjustbox}

\begin{document}


\title{Search for Dark Matter Axions with Tunable TM$_{020}$ Mode}

\author{Sungjae Bae}
\thanks{These authors contributed equally to this work.}
\affiliation{Department of Physics, KAIST, Daejeon 34141, Republic of Korea}
\affiliation{Center for Axion and Precision Physics Research, IBS, Daejeon 34051, Republic of Korea}
\author{Junu Jeong}
\thanks{These authors contributed equally to this work.}
\affiliation{Center for Axion and Precision Physics Research, IBS, Daejeon 34051, Republic of Korea}
\author{Younggeun Kim}
\affiliation{Center for Axion and Precision Physics Research, IBS, Daejeon 34051, Republic of Korea}
\author{SungWoo Youn}
\email{swyoun@ibs.re.kr}
\affiliation{Center for Axion and Precision Physics Research, IBS, Daejeon 34051, Republic of Korea}%
\author{Heejun Park}
\affiliation{Center for Axion and Precision Physics Research, IBS, Daejeon 34051, Republic of Korea}
\author{Taehyeon Seong}
\affiliation{Center for Axion and Precision Physics Research, IBS, Daejeon 34051, Republic of Korea}
\author{Seongjeong Oh}
\affiliation{Center for Axion and Precision Physics Research, IBS, Daejeon 34051, Republic of Korea}
\author{Yannis K. Semertzidis}
\affiliation{Center for Axion and Precision Physics Research, IBS, Daejeon 34051, Republic of Korea}
\affiliation{Department of Physics, KAIST, Daejeon 34141, Republic of Korea}

\date{\today}

\begin{abstract}
Axions are hypothesized particles believed to potentially resolve two major puzzles in modern physics: the strong CP problem and the nature of dark matter.
Cavity-based axion haloscopes represent the most sensitive tools for probing their theoretically favored couplings to photons in the microelectronvolt range.
However, as the search mass (or frequency) increases, the detection efficiency decreases, largely due to a decrease in cavity volume. 
Despite the potential of higher-order resonant modes to preserve experimental volume, their practical application in searches has been limited by the challenge of maintaining a high form factor over a reasonably wide search bandwidth.
We introduce an innovative tuning method that uses the unique properties of auxetic materials, designed to effectively tune higher modes.
This approach was applied to the TM$_{020}$ mode for a dark matter axion search exploring a mass range from 21.38 to 21.79\,\unit{\micro\eV}, resulting in the establishment of new exclusion limits for axion-photon coupling greater than approximately $10^{-13}$\,\unit{\per\giga\eV}.
These findings signify a breakthrough, demonstrating that our tuning mechanism facilitates the practical utilization of higher-order modes for cavity haloscope searches.
\end{abstract}

\maketitle

The Peccei-Quinn mechanism offers a compelling solution to the longstanding {\it CP} symmetry problem in quantum chromodynamics by proposing a new global $U(1)$ symmetry~\cite{PecceiQuinn:PRL:1977}.
This theoretical framework predicts the existence of the axion, a pseudoscalar particle arising as a consequence of spontaneous symmetry breaking~\cite{Weinberg:PRL:1978,Wilczek:PRL:1978}.
The invisible axions, theorized by Kim-Shifman-Vainshtein-Zakharov (KSVZ)~\cite{Kim:PRL:1979,Shiftman:NPB:1980} and Dine-Fischler-Srednicki-Zhitnitsky (DFSZ)~\cite{Zhitnitsky:SJNP:1980,Dine:PLB:1981}, are expected to have masses below the electronvolt scale and to interact minimally with ordinary matter.
From a cosmological perspective, these particles could produce an abundant, non-relativistic population in the early universe, making them a compelling candidate for understanding the composition of dark matter~\cite{Preskill:PLB:1983,Abbott:PLB:1983,Dine:PLB:1983}.

Experimental searches for dark matter axions often rely on detecting axion-induced photons in a strong magnetic field~\cite{cajohare:github:2020,Yannis:SciAdv:2022}.
Cavity haloscopes, using microwave cavities, enhance signal amplitudes through resonant phenomena, increasing sensitivity for axion detection~\cite{Sikivie:PRL:1983}.
This methodology achieves its optimal efficiency when the cavity resonant frequency matches that of the axion field, resulting in the maximum conversion power, as formulated in~\cite{Kim:JCAP:2020}:
\begin{equation}
    P_{a\gamma\gamma} = \left[\frac{g_{a\gamma\gamma}^{2}\rho_{a}}{m_{a}^{2} c/\hbar^{3}} \right] \left[ \frac{\omega_{a}}{\mu_{0}} \int\left| \mathbf{B}_e \right|^{2} dV_{c} \right] C \frac{Q_{c}Q_{a}}{Q_{c} + Q_{a}},
\end{equation}
where $g_{a\gamma\gamma}$ denotes the axion-photon coupling in units of \unit{\per\giga\eV}, $\rho_{a}$ specifies the local density of dark matter axions in the solar system in units of \unit{\giga\eV\per\cm^{3}}, and $m_{a}$ is the axion mass related to its angular frequency as $\omega_{a}$ ($ \approx m_{a}c^{2}/\hbar$).
$\mathbf{B}_e$ is the external magnetic field in a medium within the cavity volume $V_{c}$ with the vacuum magnetic permeability $\mu_{0}$, and $Q_{a}$ and $Q_{c}$ are the quality factors of the axion and the cavity, respectively.
The form factor $C$ quantifies the alignment between the electric field of the cavity resonant mode ($\mathbf{E}_{r}$) and the applied magnetic field:
\begin{equation}
    C = \frac{\left| \int \mathbf{E}_{r}\cdot \mathbf{B}_e \, dV_{c} \right|^{2}}{\int \varepsilon_{r} \left| \mathbf{E}_{r} \right|^{2}dV_{c} \int \left| \mathbf{B}_e \right|^{2} dV_{c}},
\end{equation}
where $\varepsilon_{r}$ is the dielectric constant inside the cavity.

Given that the axion mass is  unknown {\it a priori}, it is essential to scan a broad range of potential masses as quickly as possible.
An experimental parameter that characterized the search performance is the scan rate, defined as
\begin{equation}
    \frac{df}{dt} \approx \eta \left(\frac{1}{\rm SNR} \right)^{2}\left( \frac{P_{\rm sig}}{k_{B} T_{\rm sys}} \right)^{2} \frac{Q_{a}}{Q_{l}},
\end{equation}
where $\eta$ is the acquisition efficiency and $\rm SNR$ is the desired signal-to-noise ratio.
$P_{\rm sig}$ denotes the signal power extracted from the cavity through an antenna loaded with a coupling parameter $\beta$, resulting in the loaded quality factor $Q_{l}=Q_{c}/(1 + \beta)$.
$T_{\rm sys}$ represents the equivalent temperature of system noise, consisting of inherent thermal noise governed by the Johnson-Nyquist theorem~\cite{Johnson:PR:1928,Nyquist:PR:1928} and additional noise from the receiver chain.

For cylindrical geometries, the TM$_{010}$ resonant mode is favored, yet its frequency, being radius-dependent, imposes limitations due to volume reduction at higher frequencies.
To efficiently expand the search range, various resonator designs have been proposed~\cite{Hagmann1990RSI,JEONG2018412,Lawson2019PRL,Simanovskaia2021RSI,Kuo2021JCAP,McAllister2024PRD}, including the adaption of higher-order modes~\cite{Caldwell2017PRL,McAllister2018PRA,Kim2020JPG,Quiskamp2020PRA,Alesini2022PRD,Cervantes2022PRL}.
An intriguing tuning mechanism that relies on the auxetic material properties, which thicker (thinner) when stretched (compressed), has been demonstrated to achieve a considerable tuning range while preserving an adequate form factor for higher-order modes~\cite{Bae2023PRD}.
Here, we employ the auxetic structure into a hexagonal configuration to offer a viable mechanism for tuning the TM$_{020}$ mode.
This tuning mechanism was implemented to an experimental search for dark matter axions, setting new exclusion limits for axion-photon coupling over a mass (frequency) range of 21.38 (5.17) to 21.79\,\unit{\micro eV} (5.27\,\unit{\giga\Hz}).
This Letter provides a comprehensive description of the tuning mechanism and the experiment conducted.

\begin{figure}
    \centering
    \includegraphics[width=\linewidth]{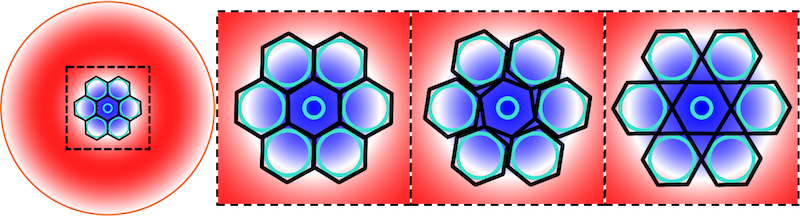}
    \caption{Leftmost: Simulated electric field distribution for the TM$_{020}$ mode in a cylindrical cavity, with red and blue indicating opposite field directions.
    Remaining: Zoomed-in views for the dashed area for various dielectric rod configurations (cyan circles), determined by the rotation of the auxetic tuning structure (black hexagons).}
    \label{fig:cavity}
\end{figure}

Higher-order resonant modes present advantages for axion searches at high frequencies compared to the fundamental mode, extending search range without volume sacrifice and providing higher quality factors.
However, they are often discarded because of notable reductions in form factor and challenges in frequency tuning. 
A periodic dielectric structure strategically placed inside a cavity has been suggested to improve form factors, allowing reasonable sensitivities~\cite{McAllister:2018,Alesini:2021}.
For the TM$_{020}$ mode of a cylindrical cavity, for example, a dielectric structure at the core of the cavity suppresses the oppositely oscillating electric field component, increasing the quantity to approximately 0.5.
The resonant frequency changes with the size of the inserted dielectric material~\cite{Kim2020JPG}.
A recent study~\cite{Bae2023PRD} utilized the unique properties of auxetic materials to tune photonic crystal cavities by altering the lattice constant of the metamaterials in two dimensions.
This versatile approach can be employed to adjust the effective size of a dielectric, offering a means for tuning the TM$_{020}$ mode.
Our engineered design features an array of small dielectric rods positioned at the cavity center.
These rods are incorporated with an auxetic structure that adjusts their spacing, effectively altering the size of the dielectric assembly, as illustrated in Fig.~\ref{fig:cavity}.

\begin{figure}
    \centering
    \includegraphics[width=\linewidth]{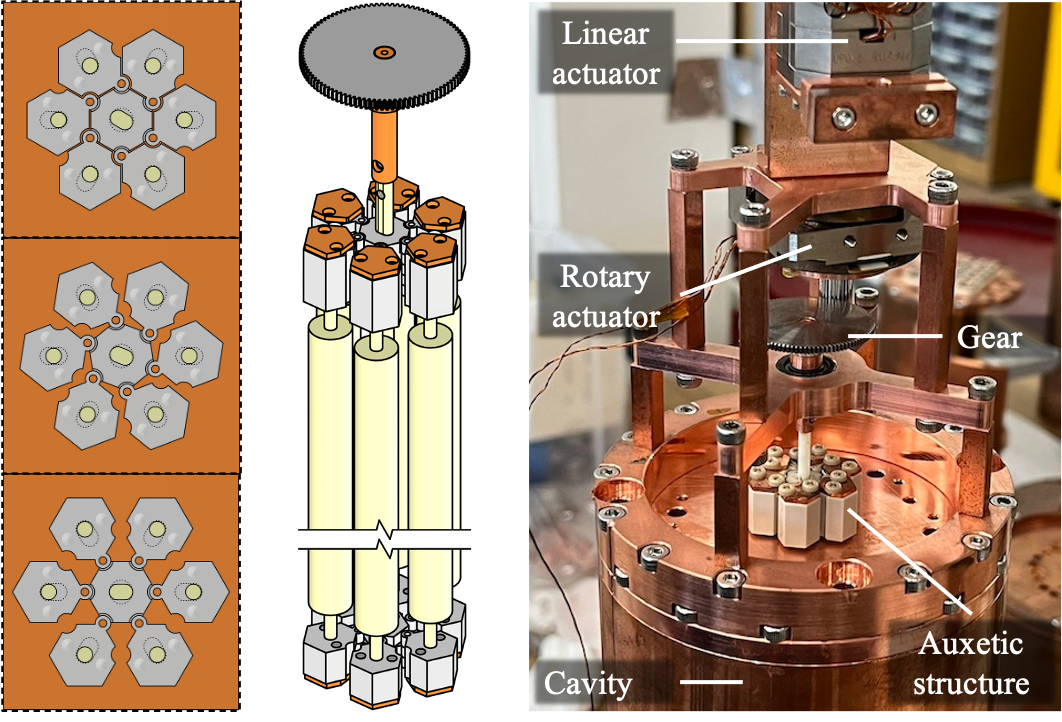}
    \caption{Left: 2-D illustration of the tuning mechanism. The rotation of the central piece drives the expansion of the auxetic structure, resulting in the configurations shown in Fig.~\ref{fig:cavity}. The dotted ovals represent the slits in the cavity endcap.
    Middle: Tuning system consisting of a central and six side dielectric rods connected to a pair of auxetic structures at the top and bottom. The central rod is integrated with a rotator via a gear system.
    Right: Photograph of the upper section of the assembled cylindrical cavity, integrating the tuning system.}
    \label{fig:hexTM020_demo}
\end{figure}

To demonstrate the tuning mechanism, a cylindrical cavity was made from oxygen-free high thermal conductivity copper with internal dimensions of $\O 78\,{\unit{\mm}} \times 300\,{\unit{\mm}}$, which yielded a detection volume of 1.43\,\unit{\L}.
The tuning structure within the cavity comprises a central rod, 3\,mm in diameter, surrounded by six side rods, each 7\,\unit{\mm} in diameter, all made of alumina.
Both ends of the side rods were machined into 2-\unit{\mm} diameter tips, which extended through slits in the cavity endcaps and connected to a pair of auxetic structures located on and beneath the cavity.
The auxetic tuning structure includes a central hexagonal piece, with each corner connected by a hinge to one of six individual pieces, all fabricated from Polyether ether ketone.
The central piece secures the central rod, while the others hold the side rods.
By creating radial slits in the cavity endcaps, the surrounding rods can move radially as the central hexagon rotates, adjusting the spacing between the rods at equal intervals.
This hexagonal auxetic structure can adjust the distance between them from 8.00\,\unit{\mm} to 9.24\,\unit{\mm}.
A finite element method simulation~\cite{multiphysics1998introduction} validated that this design maintains a form factor of approximately 0.5, more than three times that of the TM$_{020}$ mode, over a frequency range of 5.10 to 5.32\,GHz, nearly twice as high as the fundamental TM$_{010}$ mode.
A dedicated bead perturbation measurement was performed to support the simulation results~\cite{supp_mat_beadpull}.
A piezoelectric actuator attached to the central piece drives the rotation.
Additionally, to improve tuning precision and compensate for the reduction in piezoelectric constant at cryogenic temperatures, a gear system with a ratio of 100:24 was integrated with the piezo-rotator.
Figure~\ref{fig:hexTM020_demo} shows the structure of the tuning system and a photograph of the cavity assembly.
The tuning system was tested at 4.2\,\unit{\K}, successfully tuning the TM$_{020}$ resonant frequency within the range predicted by simulation.
The unloaded quality factors were measured using a vector network analyzer (VNA) through transmission, yielding approximately 110,000 across the entire tuning range.

We employed this cavity assembly for an experimental search for axion dark matter in a frequency range of 5.17--5.27\,\unit{\giga\Hz}, free of significant mode mixings.
The experiment utilized the same setup described in Ref.~\cite{Junu2020PRL}, featuring a superconducting solenoid and a wet-type He-3 cryogenic system. 
The NbTi wire solenoid generates a central magnetic field of 9\,\unit{\tesla} at a current of 81\,\unit{\A} and is capable of operating in persistent mode. 
The experiment was conducted in a 4-\unit{\K} environment, avoiding evaporative cooling of helium-4 to minimize cryogen consumption and simplify operational procedures.
The cavity assembly and cryogenic components were placed within a vacuum chamber submerged in liquid helium and cooled by injecting exchange gas.

\begin{figure}
    \centering
    \includegraphics[width=0.9\linewidth]{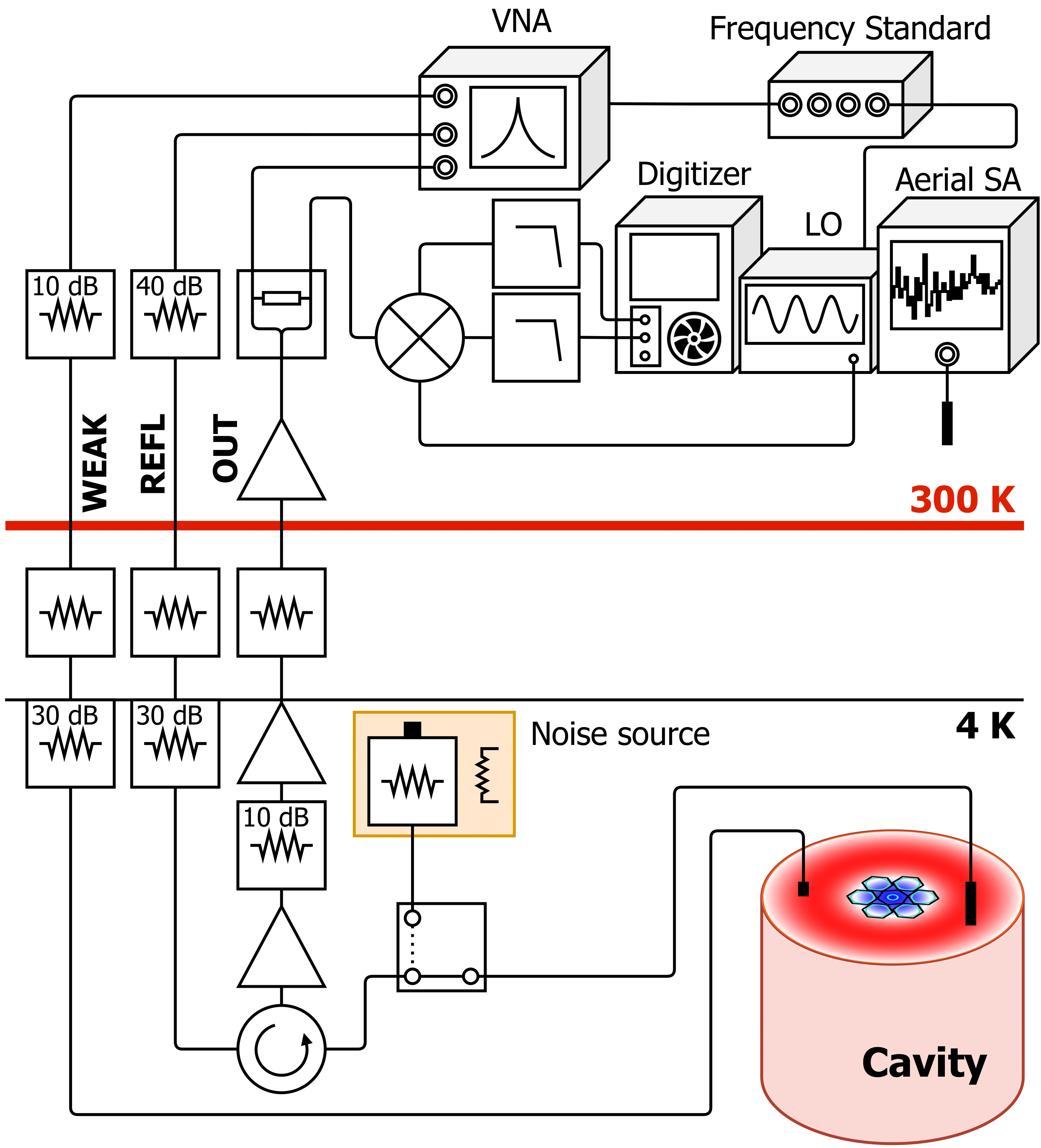}
    \caption{
    Schematic of the experimental setup illustrating the signal path from the cavity to the digitizer, VNA measurement lines, and aerial signal assessment.
    }
    \label{fig:RFchain}
\end{figure}

The experimental setup is schematized in Fig.~\ref{fig:RFchain}.
Signals captured by the strongly coupled antenna with $\beta=2$ are amplified by a series of high electron mobility transistors (HEMTs).
The amplified signals are split into two branches; one down-converted using an in-phase and quadrature (IQ) mixer for recording, while the other interfaced with a VNA port for cavity characterization via an additional antenna weakly coupled to the cavity.
An intermediate frequency (IF) of 3\,\unit{\mega\Hz} was selected based on the ambient noise spectrum.
The down-converted IQ signals are digitized at a sampling rate of 20\,\unit{\mega\Hz} and undergo software-mediated image rejection.
Additionally, an aerial spectrum analyzer (SA) is employed to detect ambient microwave signals, ensuring the integrity of internal signals.
All equipment is synchronized to a unified frequency standard for precise timing coordination.

The data acquisition (DAQ) protocol proceeded as follows. 
Initially, the rotary piezoelectric actuator tuned the TM$_{020}$ mode to the desired search frequency.
A linear piezoelectric actuator adjusted the coupling of the strongly coupled antenna to 2, and the cavity quality factor was measured using VNA transmission, yielding values around 37,000.
The local oscillator (LO) carrier frequency was set 3\,MHz lower than the search frequency and fed into the mixer for signal down-conversion.
The digitizer recorded data for 10 seconds before storing it on a computer, where a real-time discrete Fourier transform with a 100\,\unit{\Hz} resolution bandwidth converted the streaming IQ data into a power-averaged spectrum.
The cavity properties were assessed every five digitization cycles.
This process was repeated nine times, resulting in a total DAQ time of 450 seconds for each fixed search frequency.
Subsequently, the cavity was tuned to the next search frequency with an interval of approximately 50\,\unit{\kilo\Hz}, corresponding to one-third of the cavity bandwidth of about 140\,\unit{\kilo\Hz}.

The experiment ran from January 23 to February 4, 2024, yielding a total of 1638 spectra. 
An additional day was allocated for rescanning for any abnormal excesses.
Two helium transfers were conducted to replenish the evaporated liquid helium during the acquisition period, while the RF receiver chain was characterized.

\begin{figure}
    \centering
    \includegraphics[width=0.95\linewidth]{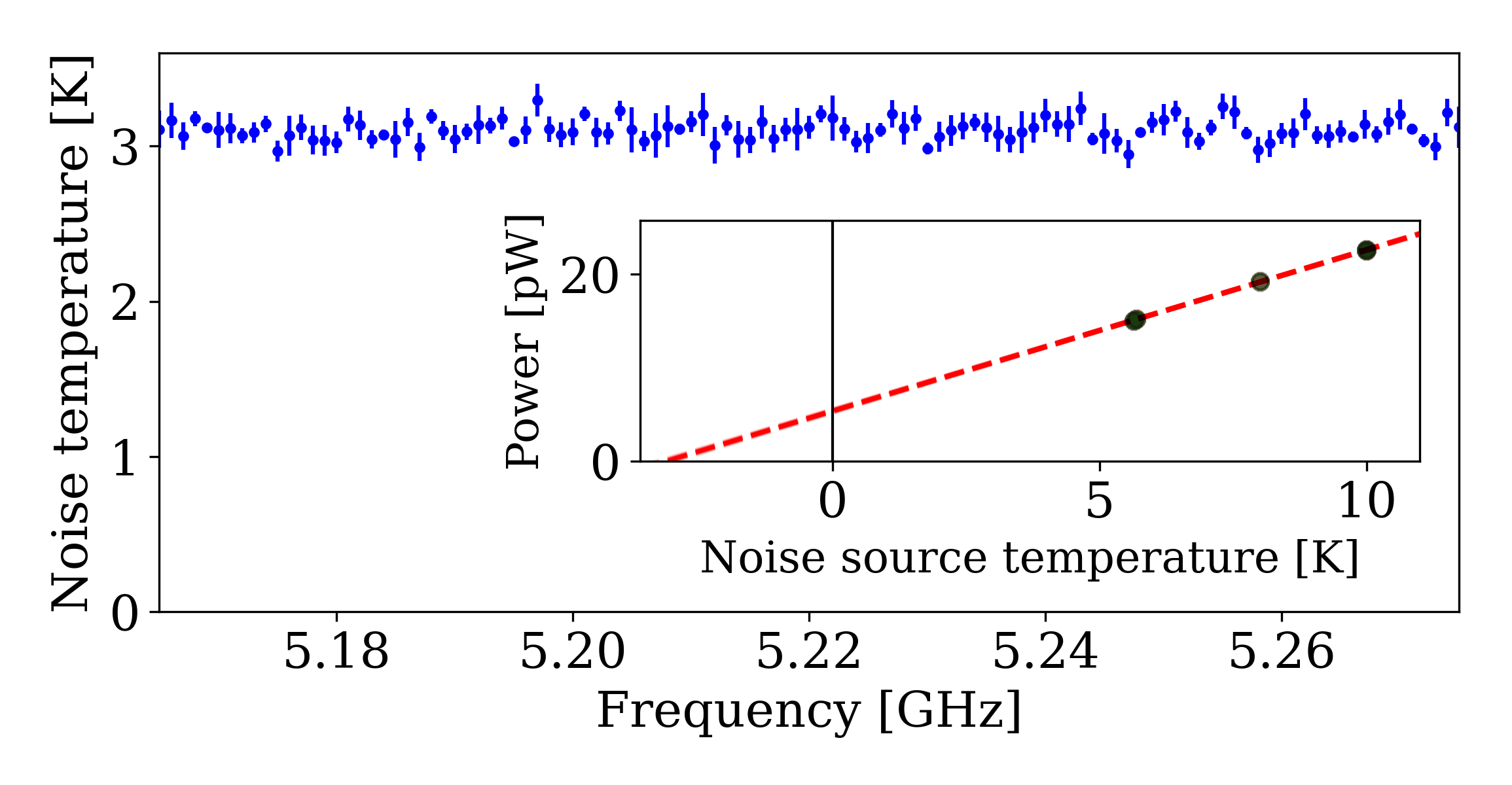}
    \caption{
    Measured noise temperature of the receiver chain using the Y-factor method.
    The inset shows the variation in power with temperature of the noise source.}
    \label{fig:Yfactor}
\end{figure}

The noise temperature of the receiver chain was measured using the Y-factor method.
An RF switch was installed between the cavity and the circulator (see Fig.~\ref{fig:RFchain}) to redirect the chain to a noise source terminated with a 50\,\unit{\ohm} load.
The temperature of the noise source was adjusted between 5\,\unit{\K} and 10\,\unit{\K} using a heater and regulated through a proportional-integral-differential controller.
Noise spectra were obtained at three different temperatures once thermal equilibrium was achieved.
Figure~\ref{fig:Yfactor} shows the measurement results, indicating a consistent noise temperature of approximately 3.1\,\unit{K} with a gain of about 71\,dB across the search frequency range.

\begin{figure*}
    \centering
    \includegraphics[width=\linewidth]{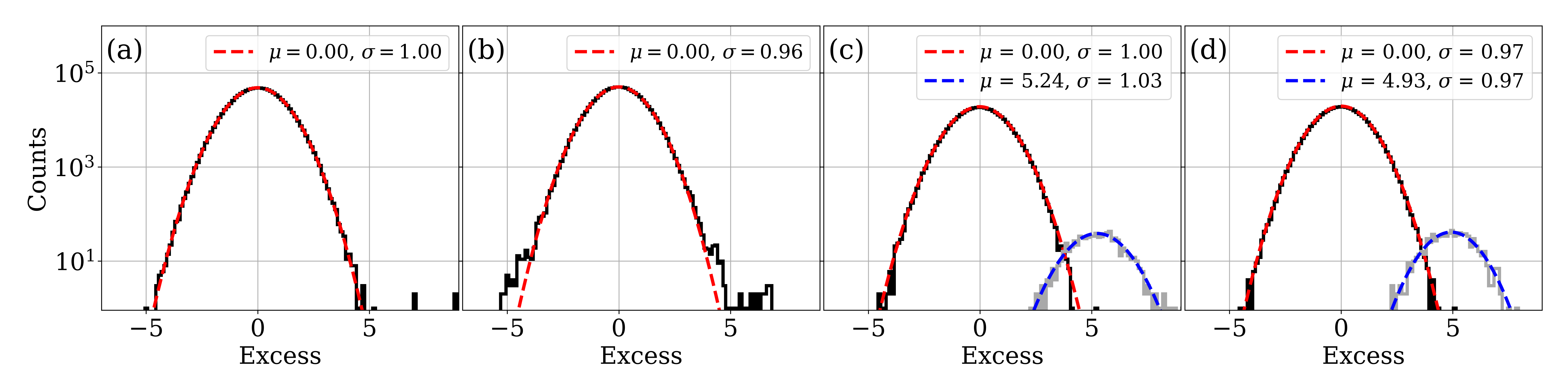}
    \caption{
    Distributions of the normalized power excesses in black for the vertically combined spectrum (a), the grand spectrum (b), and simulated grand spectra assuming a flat baseline (c) and with the baseline sampled from real data (d).
    The gray distributions in (c) and (d) represent the simulated signal samples.
    In each plot, the red (blue) dashed line in each plot represents a Gaussian fit to the noise (signal) distribution, where $\mu$ and $\sigma$ denote the mean and standard deviation, respectively. 
    }
    \label{fig:histogram}
\end{figure*}

The data analysis involved traditional processes including removal of the baseline, vertical combination, and horizontal combination~\cite{supp_mat_analysis}.
Each 10-second raw spectrum was first refined by dividing it by a fifth-order polynomial function, fitted to the sideband spectrum to correct for the frequency-dependent losses by the mixer and the filter in the IF conversion stage.
Then, a baseline was estimated within a 1\,\unit{\mega\Hz} span around the cavity frequency using a five-parameter fit, designed to account for the impedance mismatch effects of the amplifier's noise~\cite{PhysRevD.64.092003}.
For each bin, power excess was determined from the estimated baseline and then scaled according to the expected SNR, assuming KSVZ axions make up all dark matter, i.e., $\rho_a^{\rm KSVZ} = 0.45$\,\unit{\giga\eV/\cubic\cm}.
This scaling process enables expressing the final sensitivity in terms of KSVZ axion coupling.
The scaled power excess data from all tuning steps were vertically merged for corresponding frequency bins using inverse-variance weighting, creating a unified spectrum that spans the search range between 5.17 and 5.27\,\unit{\giga\Hz}.
For each frequency bin, we weighted and merged the power excesses of subsequent bins following the standard halo model for axions, characterized by a boosted Maxwell-Boltzmann distribution. 
This process across the full frequency span generates a grand spectrum, with each bin's value reflecting its maximum likelihood estimate of that frequency.
The power excess distributions of the vertical and grand spectra can be seen in Fig.~\ref{fig:histogram}(a) and (b), respectively.
The standard normal distribution of the vertical spectrum serves as a validation of the analysis procedure.
The nonunity standard deviation observed in the grand spectrum is attributed to the bin-to-bin correlation during the merging process.

The efficiency of baseline removal was estimated on the basis of Monte Carlo simulations.
Two sets of white noise were generated and added over a flat baseline and a data-driven baseline.
Each noise spectrum was overlaid with an axion signal with SNR of 5.2 and the entire analysis process was performed.
This procedure was repeated 800 times for different noise samples to collect sufficient statistics.
Examples of the corresponding power excess distributions are presented in Fig.~\ref{fig:histogram}(c) and (d), respectively.
From these, we calculated the SNR efficiency, defined as the ratio of the SNR for Fig.~\ref{fig:histogram}(d) to that for Fig.~\ref{fig:histogram}(c), obtaining an average $\epsilon_{\rm SNR}=98.0\pm2.6$\%.
The form factors were quoted from simulated values, accounting for the misalignment effect of the tuning rods in the auxetic tuning structure.
The hinges connecting the side pieces to the central piece were designed with pins 50\,\unit{\micro\metre} smaller than the holes to reduce friction.
A comprehensive simulation study assessed this effect, with each rod position randomly deviating within a uniform distribution with a width of 50\,\unit{\micro\metre}.
This procedure was repeated 1,000 times for 6 different frequencies, and the resulting form factors were interpolated across the entire frequency range.
The misalignment led to a frequency-dependent degradation of the form factor, reaching up to 1.3\% at the highest frequencies.

Following the construction of the grand spectrum, a hypothesis test was conducted to examine any abnormal excesses.
A null hypothesis of the existence of an axion with a coupling strength $\sqrt{5.2\,\sigma/\epsilon_{\rm SNR}}$ times that of the KSVZ axion was tested across all scanned frequencies.
A threshold of $3.91\,\sigma$ was set to reject the null hypothesis for frequency bins exhibiting excess below this threshold with a 90\% confidence level.
Dozens of bins failing to meet these criteria grouped into 11 clusters and flagged for rescanning.
With additional DAQ for those frequencies, all but one were confirmed to be statistical fluctuations.
The excesses that persisted after rescanning was identified as channel 36 of the wireless local area network signal, which was clearly visible over the aerial spectrum.
Consequently, the corresponding frequency bins were excluded from the search.

\begin{figure*}
    \centering
    \includegraphics[width=0.9\linewidth]{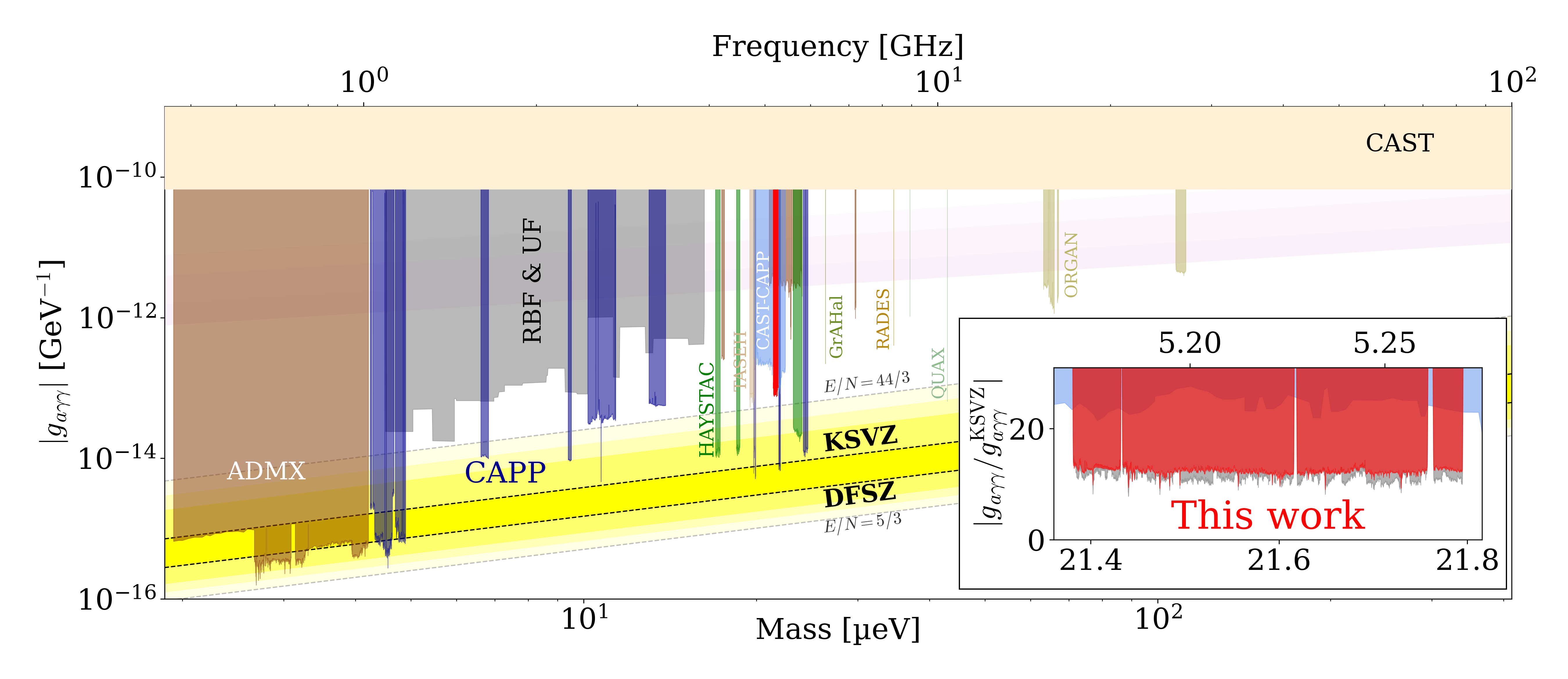}
    \caption{
    Comprehensive exclusion limits for axion-photon coupling as a function of axion mass (frequency) set by various experiments, each denoted by a distinct color and label~\cite{Junu2020PRL,PhysRevLett.59.839,PhysRevD.42.1297,PhysRevLett.80.2043,PhysRevD.64.092003,PhysRevLett.118.061302,PhysRevLett.120.151301,PhysRevLett.124.101303,PhysRevLett.124.101802,Alvarez-Melcon:2021aa,Backes:2021aa,PhysRevLett.126.191802,PhysRevLett.127.261803,Adair:2022aa,PhysRevLett.128.241805,Alesini2022PRD,PhysRevLett.129.111802,Kutlu:2022kvo,PhysRevLett.130.071002,Anastassopoulos:2017aa,PhysRevLett.133.051802,PhysRevX.14.031023}.
    The dashed black line band indicates the KSVZ~\cite{Kim:PRL:1979,Shiftman:NPB:1980} and DFSZ~\cite{Zhitnitsky:SJNP:1980,Dine:PLB:1981} models, with uncertainties represented by the yellow band, while the purple band delineates the region pertinent to ALP cogenesis~\cite{Raymond2020PRL,Raymond2021JHEP}.
    The red area corresponds to the parameter space newly excluded by this experiment.
    The inset magnifies the region with the y-axis representing the coupling strength relative to the KSVZ model in a linear scale.
    The gray shading in the inset indicates results obtained from Bayesian analysis, while the bluish shading shows the CAST-CAPP results.
    The data compilation integrates contributions sourced from~\cite{cajohare:github:2020}.}
    \label{fig:exclusion}
\end{figure*}

The exclusion limits for dark matter axions derived from the null search results are depicted in Fig.~\ref{fig:exclusion}, along with those from other experiments.
The average $\sigma$ of the grand spectrum across the entire spectrum was determined to be 27.2, leading to the rejection, at a 90\% confidence level, of axions with a coupling $g_{a\gamma\gamma}>12.0 \times g_{a\gamma\gamma}^{\rm KSVZ}$ over the mass range from 21.38 to 21.79\,\unit{\micro\eV}.
This represents approximately twice the sensitivity achieved with CAST-CAPP~\cite{Adair:2022aa} over the corresponding frequency range.
Additionally, a Bayesian analysis approach~\cite{Palken2020PRD} was applied to the same data set, and the resulting exclusion limits were illustrated as the gray area shown in the inset of Fig.~\ref{fig:exclusion}.

The following experimental uncertainties were considered in establishing the final limits.
The primary source of uncertainty came from the estimation of the noise temperature using the Y factor method, with the fitting error contributing 2.8\% to the overall uncertainty.
Simulation studies to estimate the SNR efficiency and the form factor yielded uncertainties of 2.6\% and up to 2.0\%, respectively.
Additionally, the uncertainty in the antenna coupling, estimated to be 0.7\%, was derived from the fitting error of the reflection traces.
The measured loaded quality factor, approximately 37,000, was found to have an uncertainty of $<0.1$\% when accounting for both statistical fluctuations and fitting errors from the VNA transmission traces.
Consequently, the total uncertainty affecting the axion-photon coupling, obtained from the error propagation of these factors, was determined to be 2.0\% on average.

In conclusion, we introduced a novel tuning method for higher-order resonant modes based on the auxetic behavior of metamaterial.
The mechanism was demonstrated using an auxetic structure comprising a hexagonal configuration of small dielectric rods to tune the cylindrical TM$_{020}$ mode.
Subsequently, the prototype cavity was integrated into a haloscope experiment to search for axion dark matter in the mass range between 21.38 and 21.79\,\unit{\micro\eV}.
The experimental results established a new exclusion limit for axion-photon coupling $g_{a\gamma\gamma}>12.0 \times g_{a\gamma\gamma}^{\rm KSVZ}$ ($10.9 \times g_{a\gamma\gamma}^{\rm KSVZ}$ from Bayesian analysis) across the mass range.
Our results represent an important advancement, demonstrating the feasibility of employing higher-order modes in cavity haloscope searches in high-mass regions. 
This innovation presents a more comprehensive and effective search strategy in the quest to unravel the mysteries of axion dark matter.

\begin{acknowledgments}
This work was supported by the Institute for Basic Science (IBS-R017-D1).
\end{acknowledgments}

\bibliographystyle{apsrev4-2}
\bibliography{main}

\end{document}


\title{Supplemental Material: Search for Dark Matter Axions with Tunable TM$_{020}$ Mode}

\author{Sungjae Bae}
\affiliation{Department of Physics, KAIST, Daejeon 34141, Republic of Korea}
\affiliation{Center for Axion and Precision Physics Research, IBS, Daejeon 34051, Republic of Korea}
\author{Junu Jeong}
\affiliation{Center for Axion and Precision Physics Research, IBS, Daejeon 34051, Republic of Korea}
\author{Younggeun Kim}
\affiliation{Center for Axion and Precision Physics Research, IBS, Daejeon 34051, Republic of Korea}
\author{SungWoo Youn}
\affiliation{Center for Axion and Precision Physics Research, IBS, Daejeon 34051, Republic of Korea}%
\author{Heejun Park}
\affiliation{Center for Axion and Precision Physics Research, IBS, Daejeon 34051, Republic of Korea}
\author{Taehyeon Seong}
\affiliation{Center for Axion and Precision Physics Research, IBS, Daejeon 34051, Republic of Korea}
\author{Seongjeong Oh}
\affiliation{Center for Axion and Precision Physics Research, IBS, Daejeon 34051, Republic of Korea}
\author{Yannis K. Semertzidis}
\affiliation{Center for Axion and Precision Physics Research, IBS, Daejeon 34051, Republic of Korea}
\affiliation{Department of Physics, KAIST, Daejeon 34141, Republic of Korea}

\maketitle

\date{\today}

\section{Bead perturbation}
\begin{figure}
    \centering
    \includegraphics[width=0.9\linewidth]{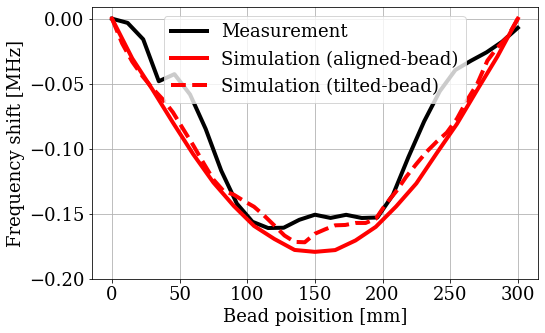}
    \caption{Bead perturbation measurement characterized by the frequency shift as a function of the bead's axial position within the cavity.
    The black and red curves correspond to the measurement data and simulated results.
    }
    \label{fig:beadpull}
\end{figure}

The field distribution for the lowest TM mode along the cavity axis was evaluated using a bead perturbation measurement~\cite{10.1063/1.5055246} conducted at room temperature, employing a small cylindrical bead ($\O 2\,{\rm mm} \times 5\,{\rm mm}$), made of alumina.
The bead was suspended with kevlar strings attached at both ends, and its position within the cavity was manually controlled. 
The resulting frequency shift as a function of the bead position is shown in Fig~\ref{fig:beadpull}. 
The black curve represents the measurement, while the red curves correspond to the simulation. 
The dashed red curve shows the simulated results when the bead is tilted (not perfectly aligned), which may more accurately reflect the actual measurement conditions. 
Although the measurement does not perfectly coincide with the simulations, the observed symmetry in the data supports a symmetric field distribution along the longitudinal direction.

\section{Data Analysis}

The data analysis method employed in this study adheres to the widely accepted procedure for haloscope searches~\cite{PhysRevLett.118.061302}.

Each raw power spectrum is down-converted by a local oscillator to an intermediate frequency of 3 MHz, then averaged, digitized, and stored every 10 seconds.
For baseline estimation, a fifth-order polynomial function is first fitted to the spectrum over a frequency range from 1\,MHz to 5\,MHz, to account for the frequency dependence of the amplifier gain and mixer conversion loss.
A 1-MHz span centered at 3\,MHz, corresponding to the cavity resonance, is excluded from the fitting.
Subsequently, a five-parameter function is applied to the spectrum over the excluded 1-MHz span to reflect the physics parameters including cavity response~\cite{PhysRevD.64.092003}, resulting in an overall baseline.
The power excess, $\delta_{p}$, for each bin is determined as the deviation of the measured power from this baseline.
The fluctuation in $\delta_{p}$, denoted as $\sigma_{p}$, is estimated from the sample standard deviation of $\delta_{p}$.
This value is confirmed by the theoretical calculation of $1/\sqrt{1000}$ for a resolution bandwidth of 100\,Hz and a power averaging time of 10 seconds.

The next step involves the so-called rescaling process, where the obtained power excess is divided by the signal-to-noise ratio (SNR) of a reference signal as
\begin{equation}
    \delta_{r} = \frac{\delta_{p}}{{\rm SNR}_{\rm ref.}} \quad {\rm and} \quad \sigma_{r} = \frac{\sigma_{p}}{{\rm SNR}_{\rm ref.}}.
\end{equation}
This process ensures that all power excesses are in consistent signal units, facilitating maximum likelihood estimation through inverse variance weighting during subsequent vertical and horizontal combinations. 
The expected signal from the KSVZ axion is used as the reference signal.

Following this, the rescaled power spectra are vertically combined. 
Vertical combination involves aggregating spectra that share the same frequency bins. 
In haloscope experiments, spectra are collected by tuning the resonator without prior knowledge of the axion frequency, resulting in multiple data points at the same frequency. 
This vertical combination is performed by summing these data points using inverse variance weighting, as follows:
\begin{equation}
    \delta_{v} = \frac{\sum \delta_{r}\sigma_{r}^{-2}}{\sum \sigma_{r}^{-2}} \quad {\rm and} \quad \sigma_{v} = \frac{1}{\sqrt{\sum \sigma_{r}^{-2}}}.
\end{equation}
This process constructs a single broad spectrum that spans the entire frequency range.

Since the signal shape of virialized axions has a finite bandwidth, combining neighboring bins within this bandwidth statistically enhances the signal.
For each frequency bin of the vertically combined spectrum, the power of the neighboring bins is weighted according to the expected axion power density distribution, $w_{a}$, and then summed up using inverse variance weighting as
\begin{equation}
    \delta_{h} = \frac{\sum \delta_{v} w_{a} \sigma_{v}^{-2}}{\sum w_{a}^{2} \sigma_{v}^{-2}} \quad {\rm and} \quad \sigma_{h} = \frac{1}{\sqrt{\sum w_{a}^{2} \sigma_{v}^{-2}}}.
\end{equation}
This process, known as horizontal combination, is repeated for all frequency bins within the scanned range. The final resulting distribution of $\delta_{h}/\sigma_{h}$ constitutes the grand spectrum.

Finally, a hypothesis test is performed for each frequency bin of the grand spectrum against a target SNR, ${\rm SNR}_{\rm target}$, typically set to 5.
By applying a threshold at the desired confidence level (CL)—for example, 3.9 for a 90\% CL—the hypothesis of the existence of signal with ${\rm SNR}_{\rm target}=5.2$ is evaluated, as described in the main text.
If the hypothesis is not rejected for certain frequency bins, additional data are collected around those bins to assess statistical fluctuations.
If the hypothesis still holds, further analysis is carried out to determine whether the signal originates from an external source or indicates the presence of dark matter. 
This may involve examining the signal's dependence on cavity resonant modes and/or magnetic field strength.

For frequency bins where the hypothesis is rejected, an exclusion limit is established.
The efficiency of the SNR, $\epsilon_{\rm SNR}$, which accounts for signal loss during the analysis, is determined through Monte Carlo simulations as detailed in the main text.
The SNR of the reference signal is given by ${\rm SNR}_{h} = \epsilon_{\rm SNR} / \sigma_{h}$.
The exclusion limit for a given coupling and dark matter density at each frequency is then calculated by scaling the ratio of ${\rm SNR}_{h}$ to ${\rm SNR}_{\rm target}$ relative to the reference values as
\begin{equation}
    \left( g_{a\gamma\gamma}^{2}\rho_{a} \right)_{\rm excl.} = \frac{{\rm SNR}_{\rm target}}{{\rm SNR}_{h}} \left( g_{a\gamma\gamma}^{2}\rho_{a} \right)_{\rm ref.}.
\end{equation}

\bibliographystyle{apsrev4-2}
\bibliography{supplementary}